\documentclass{article}
\usepackage{epsfig}

\newcommand{\fref}[1]{Figure~\ref{#1}}
\newcommand{\cref}[1]{Chapter~\ref{#1}}
\newcommand{\beq}{\begin{equation}}
\newcommand{\eeq}{\end{equation}}
\newcommand{\ba}{\begin{array}}
\newcommand{\ea}{\end{array}}
\newcommand{\bcenter}{\begin{center}}
\newcommand{\ecenter}{\end{center}}

\def\IB{\relax\hbox{$\inbar\kern-.3em{\rm B}$}}
\def\IC{\relax\hbox{$\inbar\kern-.3em{\rm C}$}}
\def\ID{\relax\hbox{$\inbar\kern-.3em{\rm D}$}}
\def\IE{\relax\hbox{$\inbar\kern-.3em{\rm E}$}}
\def\IF{\relax\hbox{$\inbar\kern-.3em{\rm F}$}}
\def\IG{\relax\hbox{$\inbar\kern-.3em{\rm G}$}}
\def\IGa{\relax\hbox{${\rm I}\kern-.18em\Gamma$}}
\def\IH{\relax{\rm I\kern-.18em H}}
\def\IK{\relax{\rm I\kern-.18em K}}
\def\IL{\relax{\rm I\kern-.18em L}}
\def\IP{\relax{\rm I\kern-.18em P}}
\def\IR{\relax{\rm I\kern-.18em R}}
\def\IZ{\relax\ifmmode\mathchoice
{\hbox{\cmss Z\kern-.4em Z}}{\hbox{\cmss Z\kern-.4em Z}}
{\lower.9pt\hbox{\cmsss Z\kern-.4em Z}}
{\lower1.2pt\hbox{\cmsss Z\kern-.4em Z}}\else{\cmss Z\kern-.4em Z}\fi}
\def\II{\relax{\rm I\kern-.18em I}}

\def\sCC{{\kern 0.27em\vrule height1.45ex width0.03em depth0em
          \kern-0.30em\rm C}}
\def\C{{\mathchoice
  {\sCC}
  {\sCC}
  {\kern 0.225em \vrule height1.05ex width0.025em depth0em \kern-0.25em \rm C}
  {\kern 0.180em \vrule height0.78ex width0.02em depth0em \kern-0.2em \rm C}
        }}
\def\sHH{{\rm I\kern-.16em{}H}}
\def\H{{\mathchoice
  {\sHH}
  {\sHH}
  {\rm I\kern-.13em{}H}
  {\rm I\kern-.13em{}H} }}
\def\sNN{{\rm I\kern-.16em{}N}}
\def\N{{\mathchoice
  {\sNN}
  {\sNN}
  {\rm I\kern-.12em{}N}
  {\rm I\kern-.10em{}N} }}
\def\sPP{{\rm I\kern-.16em{}P}}
\def\P{{\mathchoice
  {\sPP}
  {\sPP}
  {\rm I\kern-.12em{}P}
  {\rm I\kern-.10em{}P} }}
\def\sQQ{{\kern 0.27em \vrule height1.45ex width0.03em depth0em
          \kern-0.30em \rm Q}}
\def\Q{{\mathchoice
        {\sQQ}
        {\sQQ}
  {\kern 0.225em \vrule height1.05ex width0.025em depth0em \kern-0.25em \rm Q}
  {\kern 0.180em \vrule height0.78ex width0.020em depth0em \kern-0.20em \rm Q}
        }}
\def\sRR{{\rm I\kern-0.16em{}R}}
\def\R{{\mathchoice
  {\sRR}
  {\sRR}
  {\rm I\kern-0.12em{}R}
  {\rm I\kern-0.10em{}R} }}
\def\sZZ{{\rm Z\kern-0.32em{}Z}}
\def\Z{{\mathchoice
  {\sZZ}
  {\sZZ} 
  {\rm Z\kern-0.3em{}Z}     
  {\rm Z\kern-0.25em{}Z} }}  
\def\ZZZ{{\rm Z\kern-0.24em{}Z}}
\def\sII{{\rm I\kern-0.16em{}I}}
\def\I{{\mathchoice
  {\sII}
  {\sII}
  {\rm I\kern-0.12em{}I}
  {\rm I\kern-0.10em{}I} }}

\def\inbar{\,\vrule height1.5ex width.4pt depth0pt}
\font\cmss=cmss10 \font\cmsss=cmss10 at 7pt

\def\smiley{\hbox{\large$\bigcirc$\hspace{-0.80em}\raise.2ex
\hbox{$\cdot\cdot$}\kern-.61em\lower.2ex\hbox{\scriptsize$\smile$}}\ }
\def\frowny{\hbox{\large$\bigcirc$\hspace{-0.80em}\raise.2ex
\hbox{$\cdot\cdot$}\kern-.635em\lower.2ex\hbox{\scriptsize$\frown$}}\ }

\def\I{{\rlap{1} \hskip 1.6pt \hbox{1}}}

\makeatletter
\let\hangafter\@hangfrom
\makeatother

\usepackage{fortschritte}
\def\beq{\begin{equation}}                     %
\def\eeq{\end{equation}}                       %
\def\bea{\begin{eqnarray}}                     
\def\eea{\end{eqnarray}}                       
\begin {document}

\rightline{MIT-CTP-3338} 

\def\titleline{
%
%
%
Toric Duality, Seiberg Duality and Picard-Lefschetz \\Transformations                               
}
\def\email_speaker{
{\tt
%
%
          hanany@mit.edu             
}}
\def\authors{
%
%
%
%
%
Sebasti\'{a}n Franco and  Amihay Hanany           
}
\def\addresses{
%
%
%
%
Center for Theoretical Physics,
\\ Massachusetts Institute of Technology,\\
Cambridge, MA 02139, USA.\\

}
\def\abstracttext{
%
%
Toric Duality arises as an ambiguity in computing the quiver gauge
theory living on a D3-brane which probes a toric singularity. It
is reviewed how, in simple cases Toric Duality is Seiberg Duality.
The set of all Seiberg Dualities on a single node in the quiver
forms a group which is contained in a larger group given by a set
of Picard-Lefschetz transformations. This leads to elements in the
group (sometimes called fractional Seiberg Duals) which are not
Seiberg Duality on a single node, thus providing a new set of
gauge theories which flow to the same universality class in the
Infra Red.

} \large \makefront

\section{Introduction}

In \cite{toric}, it was realized that the gauge theory living on
the world volume of a D3-brane probing a toric singularity is
sometimes non-uniquely determined. Thus, by considering  D3-branes
probing non-compact, toric, singular Calabi-Yau manifolds we are
lead to more than one gauge theory with the same toric geometry as
its moduli space. This phenomenon is the essence of Toric Duality.
It is a full equivalence between distinct ${\cal N}=1$, $d=4$
gauge theories in the IR limit. Microscopic theories with
different matter content and interactions become indistinguishable
when we consider the long distance physics they describe. This
short note is a summary of recent talks given by the authors which
describes the main features of this phenomenon as well as
describing a formalism to conveniently compute various
features of gauge theories of branes on a class of singular
manifolds \footnote{Similar ideas have been used recently in the construction
of phenomenological models by wrapping D6-branes on compact, intersecting
3-cycles of Calabi-Yau manifolds \cite{uranga_local,blumenhagen}.}.

The organization of this note is as follows. In Section
\ref{section_toric_duality}, we present some examples of toric
dual theories. Section \ref{section_webs} gives a brief
introduction to $(p,q)$ webs, which are useful in defining 5d
fixed points as well as studying dynamics of 5d gauge theories but
also in describing toric varieties and their associated 4d gauge
theories. In Section \ref{section_4d}, we explain how local mirror
symmetry enables the computation of the gauge theory on the world
volume of a D3-brane probing a toric singularity using the
geometric information encoded in a $(p,q)$ web. We also exemplify
how the $(p,q)$ web machinery can be used to derive toric duals.
Based on the mirror Type IIA picture, we show in Section
\ref{section_PL} how Picard-Lefschetz monodromy transformations
point toward generalizations of Seiberg duality. Section
\ref{section_diophantine} describes the construction of invariants
for the singularities under study that generate Diophantine
equations encoding the whole set of dual theories.

\section{Toric duality}

\label{section_toric_duality}

The determination of the gauge theory on a D3-brane probing a
toric singularity was systematized in \cite{toric,phases}, by the
development of the {\it Inverse Algorithm}. This
procedure is based on the realization of toric varieties as
partial resolutions of Abelian orbifolds. Using this technique,
the cases of cones over the Zeroth Hirzebruch and toric del Pezzo
surfaces have been extensively studied. These are toric
singularities with a shrinking compact 4-cycle.

\begin{figure}[h]
  \epsfxsize = 8cm
  \centerline{\epsfbox{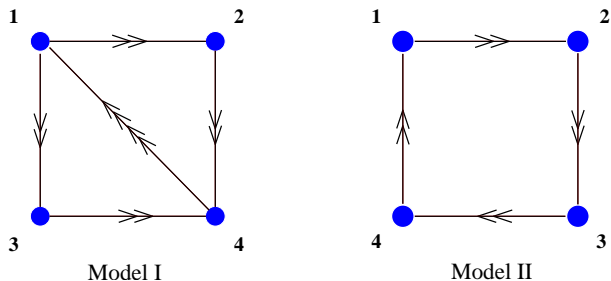}}
  \caption{Quivers for the two phases of $F_0$. Nodes represent
  $U(N)$ gauge theories, with $N$ the number of D3 branes. Each arrow
  represents a bi-fundamental field transforming under the two gauge groups
  associated to the nodes it connects.}
  \label{quivers_F0}
\end{figure}

In various cases, the resulting gauge theory is non-unique. Let us
consider the example of the Zeroth Hirzebruch surface ($F_0$) for
which the corresponding theories are displayed in
\fref{quivers_F0}.

Accompanying the different matter contents summarized in their
quivers, these models also display very distinct interactions.
These are given by the following superpotentials, which correspond
to a sum over a subset of all closed polygons in the quiver

\begin{eqnarray}
W_I & = & \epsilon_{ij} \epsilon_{mn} X_{12}^i X_{23}^m X_{31}^{jn}-
    \epsilon_{ij} \epsilon_{mn} X_{41}^i X_{23}^m X_{31}^{jn} \\ \nonumber
W_{II}& = & \epsilon_{ij} \epsilon_{mn} X_{12}^i X_{23}^m X_{34}^j
X_{41}^n \label{W_F0}
\end{eqnarray}

Finding the right subset is a difficult task in general and
presents a technical challenge. An overall re-scaling of the gauge
group ranks by a common factor $N$, as well as tracing over gauge
indices is understood in the previous quivers and superpotentials.

\section{$(p,q)$ webs}

\label{section_webs}

Five dimensional gauge theories can be engineered by 5-brane
webs in type IIB string theory \cite{webs1,webs2}. In these constructions,
the 5d gauge theories live on the $4+1$ common dimensions of the branes. The
non-trivial intersections of the branes take place on a 2-dimensional transverse plane.
The 5d theories are fully determined by the structure of the webs on this $(x,y)$ plane.

Every brane has an associated $(p,q)$ charge that dictates its tension

\beq
T_{p,q}=|p+\tau q|T_{D_5}
\label{tension}
\eeq
and its slope on the $(x,y)$ plane

\beq
\Delta x: \Delta y=p:q
\label{slope}
\eeq
where $T_{D5}$ is the D5-brane tension and $\tau$ is the complex scalar
of type IIB (which we have set to be equal to $i$ in \ref{slope}). Condition
\ref{slope} determines that 8 supercharges are preserved, leading to ${\cal N}=1$
in five dimensions.
Furthermore, $(p,q)$ charge has to be conserved at each brane intersection

\beq
\sum_i p_i=\sum_i q_i=0
\label{conservation}
\eeq

The reader is referred to \cite{webs1,webs2} for a detailed
discussion of $(p,q)$ webs, their use in engineering five
dimensional theories and explicit examples. Gauge couplings,
masses of gauge bosons and quarks, BPS spectrum and monopole
tension can be computed straightforwardly from the geometry of the
$(p,q)$ web \cite{webs1,webs2}.

Alternatively, $(p,q)$ webs can be viewed as toric skeletons defining
toric varieties \cite{LV} (see also \cite{FH} for applications of this idea
along the lines that will be discussed in this note). In this interpretation,
5-branes correspond to the loci of points at which some 1-cycles of the $T^2$
fibrations of the toric varieties shrink to zero radius.

\section{4d theories via local mirror symmetry}

\label{section_4d}

Each factor in the product gauge group of the theory on the
D-brane world-volume is given by a fractional brane. These are
bound states of D3, D5 and D7-branes, sharing four non-compact
dimensions. D3-branes are located at points (i.e. 0-cycles) on the
Calabi-Yau, while D5 and D7-branes wrap compact 2 and 4 cycles
respectively.

The corresponding quiver can be obtained by looking at the mirror
Type IIA geometry, in which D3-branes transverse to the original
non-compact Calabi-Yau map to D6-branes wrapping a $T^3$ \cite{HI}.
From the homology class of the $T^3$

\beq
[T^3]=\sum_{i=1}^n n_i S_i \ \ \ \ \ n_i \in \IZ
\eeq
we can compute the gauge group and matter content of the ${\cal N}=1$, $d=4$ gauge theory produced by a wrapped D6-brane

\beq
G=\prod_{i=1}^n U(n_i) \ \ \ \ \ I_{ij}=^\# (S_i.S_j)
\eeq

Each 3-cycle $S_i$ wraps a 1-cycle $C_i$ of a smooth elliptic fiber that degenerates at a point $z_i$. The intersection
numbers between the $S_i$'s are thus equal to the ones of the $C_i$'s which can be easily computed from their $(p_i,q_i)$
charges

\beq
^\# (S_i.S_j)= {}^\#(C_i.C_j)=\det \left( \begin{array}{cc} p_i & q_i \\ p_j & q_j \end{array} \right)
\label{intersection}
\eeq

\begin{figure}[h]
  \epsfxsize = 9cm
  \centerline{\epsfbox{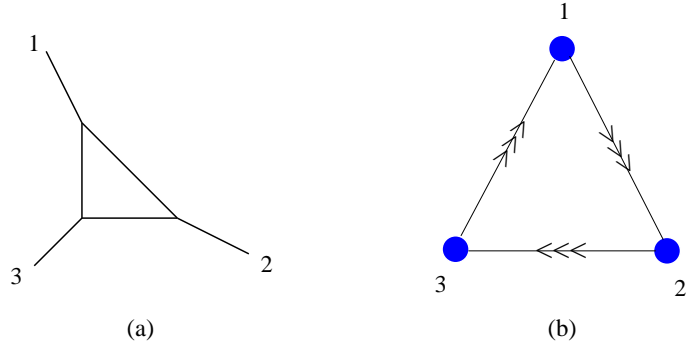}}
  \caption{(a) $(p.q)$ web and (b) quiver diagram for $dP_0$.}
  \label{dP0}
\end{figure}

Let us study how these concepts come together in the explicit
example of $dP_0$. The corresponding $(p,q)$ web is presented in
\fref{dP0}.a, from where we read the following $(p,q)$ charges

\beq
\begin{array}{ccccc}
(p_1,q_1)=(-1,2) & \ \ \ & (p_2,q_2)=(2,-1)  & \ \ \ & (p_3,q_3)=(-1,-1)
\end{array}
\eeq

Using \ref{intersection}, we compute the following intersection numbers

\beq
\begin{array}{ccccc}
^\# (C_1.C_2)=-3 & \ \ \ \ & ^\# (C_2.C_3)=-3 & \ \ \ \ & ^\# (C_3.C_1)=-3
\end{array}
\eeq
which can be conveniently summarized in the quiver diagram presented in Figure \ref{dP0}.b.

It is interesting to note that in this case, the $SU(3)$ isometry
of $\IP^2$ appears as a flavor symmetry in the gauge theory, the
$X_{ij}$ chiral fields transforming in the fundamental
representation for each pair of indices $(i,j)$. Invariance under
this $SU(3)$ fixes the superpotential uniquely

\beq W=\epsilon_{\alpha\beta\gamma} X^{(\alpha)}_{12}
X^{(\beta)}_{23} X^{(\gamma)}_{31} \eeq which is the singlet in
$X_{12} X_{23} X_{31}=3 \otimes 3 \otimes 3=1 \oplus 8 \oplus 8
\oplus 10$. This superpotential is also invariant under the
$\IZ_3$ cyclic permutations of the nodes $(123)$. Let us
demonstrate how one can compute the moduli space of vacua for this
model and reproduce the right manifold we started with. The set of
gauge invariant operators is given by 27 invariants,
$a^{\alpha\beta\gamma}=X^{(\alpha)}_{12} X^{(\beta)}_{23}
X^{(\gamma)}_{31}$. Using the F-term equations we find that any
antisymmetric combination of the indices vanishes. Therefore
$a^{\alpha\beta\gamma}$ is in the completely symmetric 10
dimensional representation of $SU(3)$. Furthermore, due to this
symmetry we find the set of equations \beq
(a^{\alpha\beta\gamma})^3=a^{\alpha\alpha\alpha}a^{\beta\beta\beta}a^{\gamma\gamma\gamma},
\eeq which is a set of 7 equations for 10 variables. A quick
inspection verifies that this is the set of equations for the
orbifold $C^3/Z_3$, the manifold we started with.

\begin{figure}[h]
  \epsfxsize = 13.5cm
  \centerline{\epsfbox{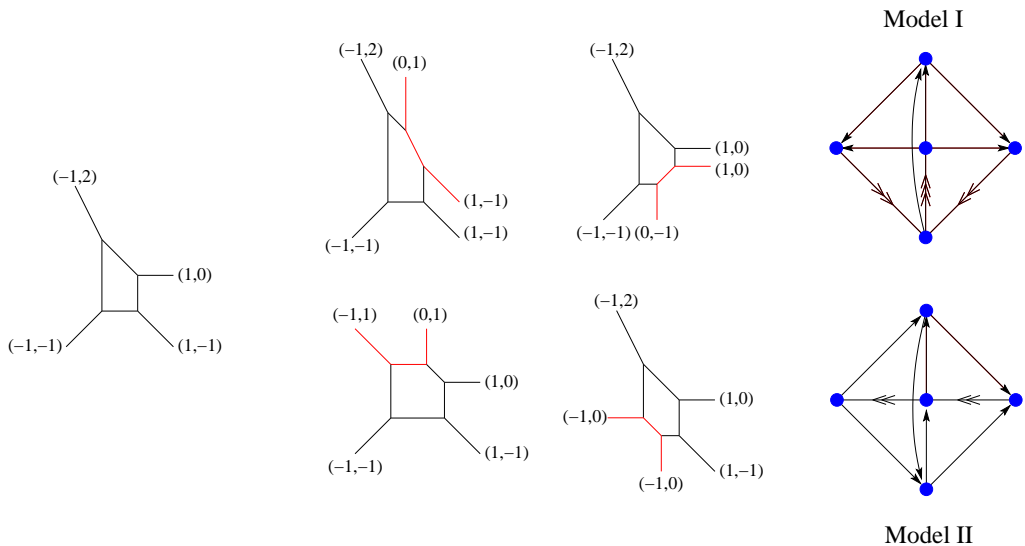}}
  \caption{Possible blowups of $dP_1$. They correspond to two inequivalent phases of $dP_2$.}
  \label{dP2}
\end{figure}

In fact, $(p,q)$ webs are powerful computational tools in deriving
toric dual theories \cite{FH}. Del Pezzo surfaces are constructed
by blowing-up up to eight generic points on $\IP^2$. A blow-up
corresponds to the replacement of a point by a 2-sphere. Since the
toric ($(p,q)$ web) representation of a 2-sphere is a segment, the
blow-up of a vertex of a given $(p,q)$ web corresponds to its
replacement by a segment. At the same time the external leg that
was originally attached to the blown-up vertex is replaced by a
pair of legs, whose $(p,q)$ charges are dictated by $(p,q)$ charge
conservation at the new vertices. Using the $SL(3,\IC)$ symmetry
of $\IP^2$, the positions of up to three generic points can be
mapped to vertices of the web. In this way, we see that, starting
from the $(p,q)$ web for $dP_0$ given in \fref{dP0}, we can
construct all the toric duals for del Pezzo surfaces up to $dP_3$,
by blowing-up vertices of the webs in every possible way. We apply
this technique to $dP_1$ in \fref{dP2} to obtain the two toric
phases of $dP_2$. It has been shown in \cite{dual,Chris2} that for
the Zeroth Hirzebruch and del Pezzo surfaces, Toric dual theories
are indeed Seiberg duals.

The discussion in this section, in addition to the one in Section \ref{section_webs}, allows
the translation of four
dimensional quantities and processes into five dimensional ones.
This approach was pursued in \cite{FH} to link the relocation of
blown-up points associated to Toric Duality to the crossing of
curves of marginal stability in related five dimensional theories.

\section{Picard-Lefschetz transformations}

\label{section_PL}

We have seen in Section \ref{section_4d} how to exploit local mirror symmetry to compute
four dimensional quiver theories from the intersections of 3-cycles
in the type IIA mirror picture. This geometric realization of the theories
suggests how to generalize Seiberg duality by using Picard-Lefschetz (PL) monodromy
transformations.

\begin{figure}[h]
  \epsfxsize = 11cm
  \centerline{\epsfbox{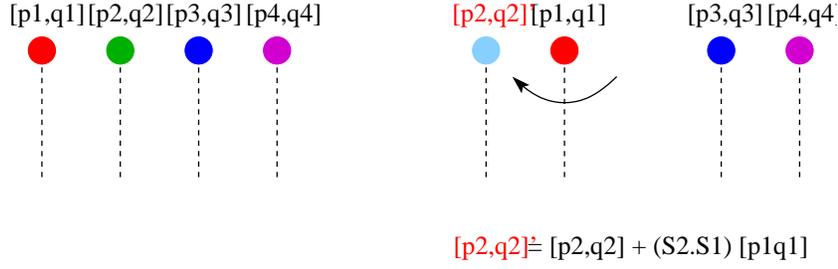}}
  \caption{Picard-Lefschetz monodromy as reordering of a configuration of 7-branes. We have
           indicated in the figure the expression of (\ref{monodromy}) as a transformation of
           $[p,q]$ charges.}
  \label{PL}
\end{figure}

PL monodromy corresponds to the reordering of vanishing cycles.
When moving a vanishing cycle $S_j$ around another one $S_i$, $S_j$ gets a contribution
proportional to $S_i$, weighted by the mutual intersection number

\beq S_j \rightarrow S_j+(S_j.S_i)S_i \label{monodromy} \eeq while
$S_i$ remains invariant. The 3-cycles $S_i$ can be represented by
$[p_i,q_i]$ 7-branes with wrapping numbers $n_i$ and we can recast
(\ref{monodromy}) in terms of $[p,q]$ charges (\fref{PL}). PL
monodromies corresponds in this language to the motion of a
7-brane across the branch cut of another one

The set of Seiberg dual theories associated to a given singularity
can be obtained by starting from a configuration of $[p,q]$
7-branes for the geometry, acting on them with PL monodromy
transformations according to the rules in (\ref{monodromy}) and
computing the resulting quiver as explained in Section
\ref{section_4d}. In fact, the group of PL transformations is
larger than the one of Seiberg dualities, and there are gauge
theories attainable with monodromies that cannot be reached by
performing any chain of Seiberg dualities on nodes of the quiver
\cite{PL}. Since further action with PL transformations results in
Seiberg duals, these theories have been named {\it Fractional
Seiberg duals}.

\begin{figure}[h]
  \epsfxsize = 13.5cm
  \centerline{\epsfbox{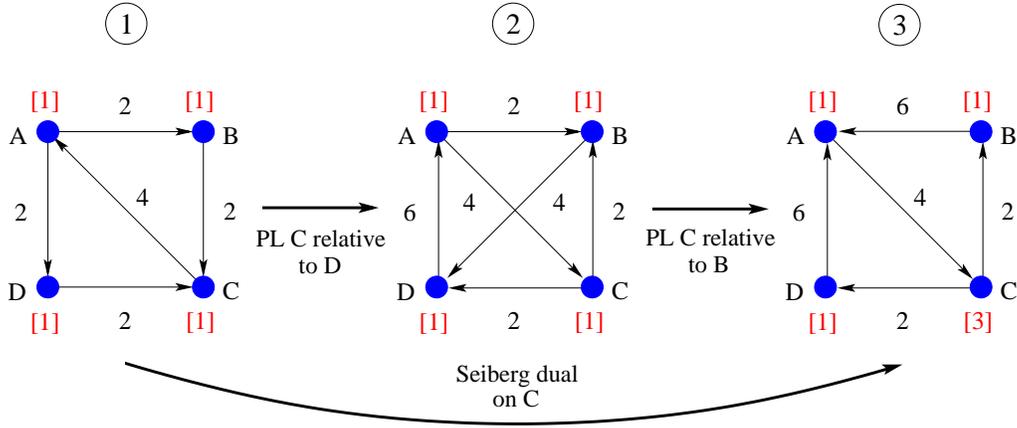}}
  \caption{A sequence of Picard-Lefschetz transformations for
  $F_0$. In this case, model 2 cannot be obtained by any combination
           of Seiberg dualities.The ranks of the gauge groups (up to an overall rescaling) are denoted in red.}
  \label{fractional_seiberg}
\end{figure}

\fref{fractional_seiberg} exhibits three theories related by PL transformations
and Seiberg dualities for the Zeroth Hirzebruch surface. In this case, Model 2 is a
new fractional Seiberg dual theory that could not have been computed by means of traditional
Seiberg duality.

\section{Diophantine equations from invariant traces}

\label{section_diophantine}

In the previous section, we have described how dual theories can be generated by performing
Picard-Lefschetz transformations on the set of degenerate fibers. All along the dualization
process there is a set of quantities that remain invariant, thus providing us with a powerful
tool in characterizing dual theories. The existence of these invariants was studied in detail
in \cite{dewolfe}. They are the number of the degenerate fibers, the greatest common divisor of the
intersection numbers and the trace of the total monodromy matrix,

\beq
K=K_{z_1} K_{z_2} ... K_{z_n}
\eeq

For the configurations of degenerate fibers under study,

\beq
Tr K=2
\eeq

This equality corresponds, for each geometry, to a Diophantine equation in the intersection
numbers that completely encodes all the gauge theories that can be generated by performing
Picard-Lefschetz transformations. As an example, let us consider the case
of $dP_0$. Using the $(p,q)$ charges in \fref{dP0} we arrive at

\beq
I_{12}^2+I_{23}^2+I_{32}^2-I_{12} I_{23} I_{32}=0
\label{diophantine_dP0}
\eeq

Since for this specific example $N_i=3 I_{jk}$, with $j,k \neq i$,
(\ref{diophantine_dP0}) can be turned into an equation for the
allowed ranks of the gauge groups

\beq
N_1^2+N_2^2+N_3^2-3 N_1 N_2 N_3=0
\eeq

{\bf Acknowledgement} We would like to thank Bo Feng, Yang-Hui He, Amer Iqbal and
Angel Uranga for collaborations in the material presented in this note. A.H.
Would also like to thank the organizers of the "35th International Symposium Ahrenshoop
on the Theory of Elementary Particles" for their hospitality. Research supported
in part by the CTP and the LNS of MIT and the U.S. Department of Energy under cooperative
agreement $\#$DE-FC02-94ER40818. A.H. is also supported by the Reed Fund Award and a
DOE OJI ward.


\begin{thebibliography}{77}

\bibitem{webs1} Ofer Aharony , Amihay Hanany , ``Branes, superpotentials and superconformal
                fixed points'', Nucl.\ Phys.\ B {\bf 504}, 239 (1997), hep-th/9704170.

\bibitem{webs2} Ofer Aharony , Amihay Hanany , Barak Kol, ``Webs of (p,q) five-branes,
                five-dimensional field theories and grid diagrams'', JHEP {\bf 9801},
                002 (1998), hep-th/9710116.

\bibitem{toric} Bo Feng, Amihay Hanany and Yang-Hui He, ``D-Brane
    Gauge Theories from Toric Singularities and Toric Duality''
    Nucl.\ Phys.\ B {\bf 595}, 165 (2001), hep-th/0003085.

\bibitem{phases} Bo Feng, Amihay Hanany and Yang-Hui He, ``Phase
    Structure of D-brane Gauge Theories and Toric Duality'',
    JHEP 0108 (2001) 040, hep-th/0104259.

\bibitem{dual} Bo Feng, Amihay Hanany, Yang-Hui He and Angel
    M. Uranga, ``Toric Duality as Seiberg Duality and Brane
    Diamonds'', JHEP {\bf 0112}, 035 (2001), hep-th/0109063.

\bibitem{Chris2} C. E. Beasley and M. R. Plesser, ``Toric Duality Is
    Seiberg Duality'', JHEP {\bf 0112}, 001 (2001), hep-th/0109053.

\bibitem{symmetries} Bo Feng, Sebastian Franco, Amihay Hanany and Yang-Hui He, ``Symmetries of toric duality'', hep-th/0205144.

\bibitem{HI} Amihay Hanany, Amer Iqbal, ``Quiver Theories from
    D6-branes via Mirror Symmetry'', JHEP {\bf 0204}, 009 (2002), hep-th/0108137.

\bibitem{LV} N.C. Leung and C. Vafa, ``Branes and Toric Geometry'',
             Adv.\ Theor.\ Math.\ Phys.\  {\bf 2}, 91 (1998), hep-th/9711013.

\bibitem{FH} Sebastian Franco and Amihay Hanany and Yang-Hui He, ``Geometric Dualities in
4d Field Theories and their 5-d Interpretation'', hep-th/0207006.

\bibitem{PL} Bo Feng, Amihay Hanany, Yang-Hui He and Amer Iqbal, ``Quiver Theories, Soliton
Spectra and Picard-Lefschetz Transformations'', hep-th/0206152.

\bibitem{dewolfe}
O.DeWolfe, T. Hauer, A. Iqbal and B. Zwiebach, ``Uncovering infinite symmetries on (p,q)
7-branes: Kac-Moody algebras  and beyond'', Adv.\ Theor.\ Math.\ Phys.\  {\bf 3}, 1835 (1999),
hep-th/9812209.

\bibitem{uranga_local}
A.~M.~Uranga, ``Local models for intersecting brane worlds'',
             JHEP {\bf 0212}, 058 (2002), hep-th/0208014.

\bibitem{blumenhagen}
R.~Blumenhagen, V.~Braun, B.~Kors and D.~Lust, ``Orientifolds of K3 and Calabi-Yau 
manifolds with intersecting D-branes'', JHEP {\bf 0207}, 026 (2002), hep-th/0206038.

\end{thebibliography}
\end{document}